\documentstyle[pre,aps,epsf,twocolumn]{revtex}
\newcommand{\bm}[1]{\mbox{\boldmath $#1$}}
\begin{document}
\draft
\title{Relaxation of finite perturbations: 
beyond the Fluctuation-Response relation}

\author{G. Boffetta$^{1}$, G. Lacorata$^{2}$, 
S. Musacchio$^{1}$ and  A. Vulpiani$^{3}$}
\address{$^{1}$Dipartimento di Fisica Generale and INFM,
Universit\`a di Torino,\\
Via Pietro Giuria 1, 10125 Torino, Italy.}
\address{$^{2}$ISAC-CNR, Sezione di Lecce, S.P. Monteroni,
         73100 Lecce, Italy}
\address{$^{3}$Dipartimento di Fisica and INFM 
(U.d.R. and SMC Center),\\ 
Universit\`a di Roma ``La Sapienza'',
P.le Aldo Moro 5, 00185 Roma, Italy}
\date{\today}

\maketitle

\begin{abstract}
We study the response of dynamical systems to finite amplitude perturbation.  
A generalized Fluctuation-Response relation is derived, 
which links the average relaxation toward equilibrium 
to the invariant measure of the system and points out the relevance  
of the amplitude of the initial perturbation. 
Numerical computations on systems with many characteristic times 
show the relevance of the above relation in realistic cases.

\end{abstract}

\pacs{PACS NUMBERS: 05.45-a, 47.52+j}


\section*{Lead Paragraph}
\label{sec:0}
{\bf 
Understanding the behavior of a dynamical system out of its 
equilibrium is a crucial issue of statistical physics.  
In the case of an infinitesimal perturbation that shifts the system 
out of equilibrium, the classical Fluctuation-Response theorem
allows to determine the linear response of the system in term 
of its equilibrium properties, i.e. correlation functions.  
While the behavior of infinitesimal perturbations 
gives relevant information for problems of statistical mechanics, 
for climate and geophysical models the main goal is to  
characterize the relaxation of large perturbations, 
which can not be obtained from the linear response theorem. 
We present here a 
generalization of Fluctuation-Response relation, 
which holds for finite amplitude perturbations, providing 
a tool for extracting non equilibrium behavior 
out of equilibrium features of the system. 
We also discuss the non trivial role of the amplitude of perturbations  
in systems where many characteristic time scales are present. 
}

\section{Introduction}
\label{sec:1}
The Fluctuation-Response (F/R) relation has a deep relevance 
in statistical physics and more generally in systems with chaotic dynamics
(in particular in hydrodynamics \cite{K59}).
The relevance of a connection between ``non equilibrium'' 
features (i.e. response to an external perturbation) 
and ``equilibrium'' properties (i.e. time correlations
computed according to the invariant measure)
is well known in statistical mechanics.
We can mention the important Green - Kubo formulas in the 
linear response theory \cite{KTH85}.
Beyond statistical physics, an other field 
where the F/R problem has an obvious relevance is climate research
\cite{L75}.
One of the key problems is the possibility to understand the response 
of the present climate to some violent changes 
(e.g. a volcanic eruption). The essential point is the 
possibility that the recovery of the climate system from a perturbation 
(response) can be estimated from its time history 
(correlations time of the unperturbed system). 

Assuming that the system is mixing and
has invariant probability density function (pdf) $\rho({\bm x})$, it is
possible to derive the following F/R relation.
Let us denote by ${\bm x}(t) = (x_1(t), \cdots, x_N(t))$ the state 
of the system at time $t$.
If at the initial time $t=0$ the system is perturbed by 
$\delta {\bm x}(0) = (\delta x_1(0), \cdots , \delta x_N(0))$,
the average evolution of the perturbation $\langle \delta x_i(t) \rangle$ 
with respect the unperturbed trajectory is
\begin{equation}
\langle \delta x_i(t) \rangle = \sum_i R_{i,j}(t) \delta x_j(0)
\label{eq:1.1}
\end{equation}
where 
\begin{equation}
R_{i,j}(t) = \langle {\delta x_i(t) \over \delta x_j(0)} \rangle = 
\langle x_i(t) f_j({\bm x(0)}) \rangle
\label{eq:1.2}
\end{equation}
and the function $f_j$ depends on $\rho({\bm x})$ as
\begin{equation}
f_j({\bm x}) = - {\partial \ln \rho({\bm x}) \over \partial x_j}
\label{eq:1.3}
\end{equation}
In Section \ref{sec:2} we will give a complete derivation of the above
formulas.

As far as we know, 
the F/R problem had been studied only for infinitesimal perturbations. 
For statistical mechanics problems it is relevant to deal with 
infinitesimal perturbations on the microscopic variables.
In a similar way this problem has importance 
in many analytical approaches to the statistical description 
of hydrodynamics where Green functions are naturally 
involved both in perturbative theory and closure schemes
\cite {K59,McC91}.

On the other hand in geophysical or climate problems 
the interest for infinitesimal perturbation
seems to be rather academic, while the interesting problem is the behavior 
of relaxation of large fluctuations in the system due to fast changes 
of the parameters.

In this paper we want to address the problem of the F/R relation 
for non infinitesimal perturbations. 
In Section \ref{sec:2} we will show that it is possible to 
generalize the F/R relation to large perturbations, involving
rare events of the invariant measure.
Section~\ref{sec:3} is devoted to a discussion on the connections,
and differences, between our approach and well known results in
dynamical system theory.
In Section \ref{sec:4} we will discuss the application to 
systems involving a single characteristic time, while Section
\ref{sec:5} is devoted to system with many characteristic times.
Section~\ref{sec:6} is devoted to conclusions and the Appendix~\ref{app:1}
contains some technical remarks.

\section{Theoretical background}
\label{sec:2}
In the following we will consider a dynamical 
system with evolution ${\bm x}(t)=\phi^{t}{\bm x}(0)$
of the $N$-dimensional vector ${\bm x}$. For generality, we 
will explicitly consider the case in which the time evolution
can also be not completely deterministic (e.g. stochastic differential
equations).
We will assume the existence of an invariant
probability distribution $\rho({\bm x})$ and the ergodicity
of the system so that
\begin{equation}
\langle A \rangle \equiv \lim_{T \to \infty} {1 \over T}
\int_{0}^{T} A({\bm x}(t)) dt = 
\mu(A) \equiv \int A({\bm x}) \rho({\bm x}) d{\bm x}
\label{eq:2.1}
\end{equation}
for any (smooth enough) observable $A$.

Our aim is the understanding of the mean response
$\langle \delta A(t) \rangle$ of a generic observable $A$ initially 
perturbed  with  $\delta A(0)$.
The first step is the study of one component of ${\bm x}$ i.e.  
$\langle \delta x_i(t) \rangle$ with an initial (non-random) perturbation
$\delta {\bm x}(0)=\delta {\bm x}_{0}$.
Introducing the probability of transition 
from $({\bm x_0},0)$ to $({\bm x},t)$, $W ({\bm x_0},0 \to {\bm x},t)$
(for a deterministic system we have
$W ({\bm x_0},0 \to {\bm x},t)=\delta({\bm x}(t)-\phi^{t}{\bm x}_{0})$),
we can easily write an expression 
for the mean value of the variable computed along the perturbed trajectory 
$x'_i(t) =x_i(t) + \delta x_i(t)$:
\begin{equation}
\left \langle x'_i(t) \right \rangle = 
\int\!\int x_i \rho' ({\bm x_0}) 
W ({\bm x_0},0 \to {\bm x},t) \, d{\bm x} \, d{\bm x_0}
\label{eq:2.3}
\end{equation}
where $\rho' ({\bm x})$ is the initial distribution of perturbed system,
which is related to the invariant distribution by 
$\rho' ({\bm x_0}) = \rho ({\bm x_0} - \delta {\bm x_0})$.
Noting that the mean value of $x_i(t)$ can be written in a similar way:
\begin{equation}
\left \langle x_i(t) \right \rangle = 
\int\!\int x_i \rho ({\bm x_0}) 
W ({\bm x_0},0 \to {\bm x},t) \, d{\bm x} \, d{\bm x_0}
\label{eq:2.2}
\end{equation}
one has:
\begin{eqnarray}
\langle \delta x_i(t) \rangle &=&
\int\!\int x_i 
{\rho ({\bm x_0} - \delta {\bm x_0}) - \rho ({\bm x_0}) \over
\rho ({\bm x_0}) }
\rho ({\bm x_0}) \nonumber \\
&\times& W ({\bm x_0},0 \to {\bm x},t) 
\, d{\bm x} \, d{\bm x_0} \nonumber \\
&=& \langle x_i(t) F({\bm x_0},\delta {\bm x}_0) \rangle
\label{eq:2.5}
\end{eqnarray}
where
\begin{equation}
F({\bm x_0},\delta {\bm x}_0) =
\left[ {\rho ({\bm x_0} - \delta {\bm x_0}) - \rho ({\bm x_0}) \over
\rho ({\bm x_0})} \right]
\label{eq:2.6}
\end{equation}

For an infinitesimal perturbation 
$\delta {\bm x}(0) = (\delta x_1(0) \cdots \delta x_N(0))$
expanding (\ref{eq:2.6}) to first order one ends with the expression
\begin{eqnarray}
\langle \delta x_i(t) \rangle &=&
- \sum_j
\langle x_i(t) \left. {\partial \ln \rho({\bm x}) \over \partial x_j} 
\right|_{t=0} \rangle \delta x_j(0) \nonumber \\
&\equiv&
\sum_j R_{i,j}(t) \delta x_j(0)
\label{eq:2.7}
\end{eqnarray}
which defines the linear response 
\begin{equation}
R_{i,j}(t) =
- \langle x_i(t) \left. {\partial \ln \rho({\bm x}) \over \partial x_j} 
\right|_{t=0} \rangle 
\label{eq:2.7b}
\end{equation} 
of the variable $x_i$ with respect to a perturbation of $x_j$.
Relation (\ref{eq:2.7b}) is 
the generalization for non Hamiltonian systems of the well known 
fluctuation/response (F/R) relation \cite{KTH85}.

Let us note that in the general case the 
invariant measure $\rho({\bm x})$ is not known, so the equation (\ref{eq:2.7b})
gives just a qualitative information. 
In the case of Gaussian distribution, $\rho({\bm x})$ factorizes and
the linear response recovers the correlator 
\begin{equation}
R_{i,j} (t) =  
\frac{\langle x_i(t) x_j(0) \rangle - \langle x_i \rangle \langle x_j \rangle}
{\langle x_j x_j \rangle - \langle x_j \rangle \langle x_j \rangle}
\label{eq:2.8}
\end{equation}

In the case of finite perturbations, the F/R relation (\ref{eq:2.5}) is 
typically non-linear in the perturbation $\delta{\bm x}_0$ and thus no
simple relations analogous to (\ref{eq:2.7b}) exist.
Nevertheless we can disentangle the different contributions
in the response (\ref{eq:2.5}) by studying an initial perturbation whose  
only non-zero component is the $j-th$ one,
\begin{equation} 
\delta^{(j)} {\bm x}(0) = (0, \cdots,0,\delta x_j(0),0,\cdots,0)
\label{eq:2.8b}
\end{equation} 
We therefore generalize the F/R relation (\ref{eq:2.7b}) 
to non-linear response of $x_i$ to a perturbation on the $j$ variable as
\begin{equation}
R_{i,j}(t)=\langle x_i(t) f_j(0) \rangle
\label{eq:2.10}
\end{equation} 
where $f_j$ is given by  
\begin{equation}
f_j({\bm x_0}) = 
{\rho ({\bm x_0} - \delta^{(j)} {\bm x}(0)) - \rho ({\bm x_0}) \over
\rho ({\bm x_0}) \delta x_j(0)}
\label{eq:2.9}
\end{equation}

The explicit prediction of the response from (\ref{eq:2.10})
requires the analytic expression of the invariant pdf, 
which is in general not known.  
Nevertheless (\ref{eq:2.5}) guarantees the existence of a
link between equilibrium properties of the system and the response 
to finite perturbations. 
This fact has a relevant consequence
for systems with one single characteristic time:     
a generic correlation (e.g. the correlation (\ref{eq:2.8})) in principle
gives informations on the relaxation time of finite size perturbations, 
even when the invariant measure $\rho$ is not known \cite{Biff02}.

\section{Remarks on the connections between F/R relation,
dynamical system theory and statistical mechanics}
\label{sec:3}
Since the F/R relation involves the evolution of differences between
variables computed on two different realizations of the system,
it is natural to conclude that this issue is closely related to
the predictability problem and, more in general, to 
chaotic behavior. Actually, a detailed analysis shows that the two
problems, i.e. F/R relation and predictability, have only a 
very weak connection. For the sake of completeness, we shortly 
discuss here the analogies and differences between these two
issues.

The typical problem for the characterization of predictability is the
evolution of the trajectory difference $\delta {\bm x}(t)$, in particular of
$\langle \ln |\delta {\bm x}(t)| \rangle$ which defines
the leading Lyapunov exponent $\lambda$.
For small $|\delta {\bm x}(0)|$ and large enough $t$ one has
\begin{equation}
\langle \ln |\delta {\bm x}(t)| \rangle \simeq 
\ln |\delta {\bm x}(0)| \lambda t
\label{eq:3.1}
\end{equation}
On the other hand, in F/R issue one deals with averages of 
quantities with sign, such as $\langle \delta {\bm x}(t) \rangle$.
This apparently marginal difference is very important and it
is at the basis of the famous objection by van Kampen
related to the standard derivation of the linear response theory
\cite{VK71}. In a nutshell, using the modern dynamical systems 
terminology, the van Kampen's argument is as follows.
Since in presence of chaos $|\delta {\bm x}(t)|$ grows 
exponentially in time, it is not possible to
linearize (\ref{eq:2.6}) for time larger than 
$(1/ \lambda) \ln (\Delta/|\delta{\bm x}(0)|)$,
where $\Delta$ is the typical fluctuation of the variable ${\bm x}$.
As a consequence, the linear response theory is expected to be valid only for
extremely small and unphysical perturbations, in clear disagreement
with the experience. A solution of this apparent paradox was proposed 
by Kubo
who suggested that ``{\it instability [of the trajectories] instead favors 
the stability of distribution functions, working as the cause of the mixing}''
\cite{kubo}.
More recent works have demonstrated the constructive role of chaos 
in F/R relation and the non relevance of van Kampen's criticism
\cite{F90,Falc95}. The objection by van Kampen remains
nevertheless relevant for numerical computations of F/R relation
(see Appendix).

Fluctuation/response relation was developed in the context of statistical
mechanics of Hamiltonian systems, but it also holds for
non conservative systems, and even non deterministic systems
(e.g. Langevin equations) and has no general relation with 
``chaotic quantities'' such as Lyapunov exponents or
Kolmogorov-Sinai entropy.
This generated in the past some confusion about the applicability
of F/R/ relation. For example, some authors claimed
(with qualitative arguments) that in fully developed 
turbulence there is no relation between equilibrium fluctuations 
and relaxation to equilibrium \cite{Rose78}
while the correct statement concerns the non validity 
of the simplified relation (\ref{eq:2.8b}) which holds only 
for systems with Gaussian statistics.

Thanks to its general validity and robustness,   
the F/R relation has been also used to obtain informations on
the unknown invariant measure $\rho ({\bm x})$ on the basis
of the linear response $R_{i,j}(t)$.
An important example comes from the field of disordered systems
where the F/R had been applied to the study of
aging phenomena \cite{Cugliandolo97}.

Concluding this short discussion on the connections 
between F/R relation, dynamical system theory 
and statistical mechanics, we mention recent results about
rigorous derivation of the Onsager reciprocity relations
\cite{Gabrielli1999} and the macroscopic fluctuation theory for
stationary non-equilibrium states \cite{Bertini2002} in
a class of stochastic models describing  interacting
particles systems.

\section{Systems with a single characteristic time}
\label{sec:4}
Let us start by studying two examples of systems with a single
characteristic time: a deterministic chaotic system (the Lorenz model)
and a nonlinear Langevin process.

We first consider the Lorenz model \cite{Lor63}
\begin{eqnarray}
{dx \over dt} & = & \sigma (y-x) \nonumber \\
{dy \over dt} & = & -xz +rx -y \\
\label{eq:4.5}
{dz \over dt} & = & xy -b z \nonumber
\end{eqnarray}
with standard parameters for chaotic behavior: $b=8/3$, $\sigma=10$ and 
$r=28$.
The correlation function (\ref{eq:2.8}) for the variable $z$,
shown in Fig.~\ref{fig1}, qualitatively reproduces the behavior
of the response to different sizes of the perturbation of the $z$ variable, 
ranging from infinitesimal ones up to the size of the attractor.
The accuracy does not increase when decreasing the perturbation 
because the invariant distribution is
not Gaussian (see Fig.~\ref{fig1}) and thus the general correlation
(\ref{eq:2.7b}) should be used.
We observe that the use of (\ref{eq:2.7b}) instead of (\ref{eq:2.8}) is
in general much more difficult because the invariant distribution
is in general non factorable.

To better illustrate this point, let us now consider
a system whose invariant probability distribution is known.
In this case we can quantitatively compare the differences 
between the responses to infinitesimal and finite perturbations.
Our example is provided by the stochastic process $x(t)$ determined by  
\begin{equation}
{dx \over dt} = - {dU(x) \over dx} + \sqrt{2D} \xi(t)
\label{eq:4.1}
\end{equation}
where $\xi(t)$ is a white noise, i.e. a Gaussian process with 
$\langle \xi(t) \rangle = 0$ and
$\langle \xi(t) \xi(t') \rangle = \delta (t-t')$.      
The invariant probability distribution is \cite{Gard}:
\begin{equation}
\rho(x) = {\mathcal N} e^{-U(x)/D}
\label{eq:4.2}
\end{equation} 
where ${\mathcal N}$ is fixed by normalization.

A Gaussian pdf is obtained using $U(x)= x^2/2$ 
which corresponds to the linear Ornstein-Uhlenbeck process 
$dx/dt = - \, x + \sqrt{2D} \xi(t)$.
Our example uses a modified version of the Gaussian case,
\begin{equation}
U = \left\{ 
\begin{array}{ll}
{1 \over 2} x^2 &, |x| < 1 \\
|x| -{1 \over 2} &, |x| > 1 .
\end{array} 
\right.
\label{eq:4.3}
\end{equation}
The resulting pdf, shown in the inset of Fig.~\ref{fig2},
has a Gaussian core, with exponential tails.
Figure~\ref{fig2} also shows the response function for
an infinitesimal and for a finite size perturbation.
For both perturbations, the response function measured from 
the perturbed trajectories is exactly predicted by statistics
of the unperturbed system according to (\ref{eq:2.10}),
while the Gaussian correlation 
$C(t) = \langle x(t)x(0) \rangle / \sigma^2$ gives only
an estimate of the relaxation time. 
By construction, the pdf of this system has larger tails than 
in the Gaussian case, thus large fluctuations decay slower than small ones.
In the linear case the mean response is simply $R(t) = \exp(- t)$
and does not depend on the amplitude of the initial perturbation 
$\delta x(0)$.

The results obtained for the Lorenz model and for the nonlinear 
Langevin equations suggest that if only one characteristic 
time is present, the existence of the F/R relation allows for some 
qualitative results even in the
absence of precise knowledge of $\rho$, both for infinitesimal and 
finite perturbation.  

\section{Systems with many characteristic times}
\label{sec:5}

In systems with many characteristic times,
different correlation functions do not show the same behavior, 
i.e. depending on the observable one can observe very different time scales,
corresponding to the different decay 
times of the correlation functions 
$C_{j,j} = \langle x_j(t) x_j(0) \rangle$ \cite{Biff02}.
In addition, at variance with systems with one single time scale, 
here the amplitude of the perturbation  
can play a major role in determining the response, 
because different amplitudes may affect features 
with different time properties.

The link between equilibrium and relaxation properties 
established by the F/R relation (\ref{eq:2.10}) 
suggests that it is possible to relate different relaxation rates with  
the time scales measured by means of correlations.   
Consider the case of an observable $A$ which depends on all 
the variables of the system $\{ x_1, \cdots, x_N \}$. 
For infinitesimal perturbations, a
straightforward generalization of (\ref{eq:1.1},\ref{eq:1.2}) gives:
\begin{equation} 
\langle \delta A(t) \rangle = \sum
\langle A \left( {\bm x}(t) \right) f_j \left( {\bm x}(0) \right) \rangle
\delta x_j(0) 
\label{eq:5.7}
\end{equation}

In the case of finite perturbations, as 
stressed in Sect. (\ref{sec:2}), it is possible to write 
a F/R relation: 
\begin{equation} 
\langle \delta A(t) \rangle = 
\langle A \left( {\bm x}(t) \right)
F \left( {\bm x}(0), \delta {\bm x}(0) \right) 
\rangle
\label{eq:5.8}
\end{equation}
in which, at variance with (\ref{eq:5.7}), all the variables are mixed.
In (\ref{eq:5.8}) the relaxation properties depend
explicitly on the initial perturbation $\delta {\bm x}(0)$.

Depending on the choice of $A({\bm x})$, different perturbations 
on $A$ correspond to different amplitudes of the perturbations 
on each variable $x_j$. Consequently, one can think that 
it is possible to associate each perturbation to a certain 
subset of variables which are mainly perturbed. 
The relaxation of $\langle \delta A(t) \rangle$ 
will be ruled by the characteristic time of that particular subset.

In order to illustrate this issue we consider a
shell model for turbulence \cite{BJPV98}.
Shell models are a simplified model for turbulent energy cascade, 
that describe the dynamics of velocity fluctuations at a 
certain scale $\ell_n = k_n^{-1}$ with a single shell-variable $u_n$. 
Wave-numbers $k_n$ are geometrically spaced as $k_n = k_0 \lambda^n$, 
allowing to cover a large range of scales with relatively few variables. 
A quadratic interaction between neighbor shell reproduces the 
main features of three-dimensional turbulence.  
The specific model we will use is
\begin{eqnarray}
\left( {d \over dt} + \nu k_n^2 \right) u_n &=&
i \left[ k_{n+1} u_{n+1}^* u_{n+2} -\epsilon k_n u_{n+1} u_{n-1}^* 
\right. \nonumber \\
&+& \left. (1-\epsilon) k_{n-1} u_{n-2} u_{n-1} \right] + f_n
\label{eq:5.1}
\end{eqnarray}
where $\nu$ is the molecular viscosity, 
$f_n$ is an external forcing which injects energy at large scale,
and $\epsilon$ is a free parameter. 
In order to have the correct conservation laws (energy and helicity) 
in the inviscid unforced case one has to fix $\epsilon = 1/2 $. 
The observable considered is the total energy 
$E(t) = {1 \over 2} \sum_{n=1}^N |u_n(t)|^2$
which is the conserved quantity in the inviscid, unforced limit
\cite{BJPV98}.

In order to study the response to perturbations with 
different amplitude on $E$,
we consider the following perturbed systems 
labeled with $i = 1,..,N$:
$u_n^{(i)}(t) = u_n(t) + \delta u_n^{(i)}(t)$ 
where the initial perturbations $\delta u_n^{(i)}(0)$
are set in the following way:   
\begin{equation}
\delta u_n^{(i)}(0) = \left\{ \begin{array}{ll}
 0 & \, , \, 1 \le n \le i-1 \\
 \sqrt{\langle |u_n|^2 \rangle} & \, , \, i \le n \le N 
 \end{array} \right.
\label{eq:5.4}
\end{equation}
This corresponds to a set of initial perturbations of the energy 
\begin{equation}
\langle \delta E_i(0) \rangle = {1 \over 2} \sum_{n=i}^N  \langle|u_n|^2 \rangle
\label{eq:5.2}
\end{equation}

Such a perturbation is motivated by the fact that 
in the unperturbed system 
the energy is distributed among the shells according to the 
Kolmogorov scaling $ \langle |u_n|^2 \rangle \sim  k_n^{-2/3}$,
and the smaller scales give smaller contributions to the energy $E(t)$. 
Thus it is natural to assume that a small perturbation of the energy
will affect mainly the small scales.

For each perturbation $\delta E_i$, 
the average response of energy
\begin{equation}
\left \langle {\delta E_i(t) \over \delta E_i(0)} \right \rangle =
\left \langle {\sum_{n=1}^N |u_n^{(i)}(t)|^2 - |u_n(t)|^2 \over 
\sum_{n=1}^N |u_n^{(i)}(0)|^2 - |u_n(0)|^2} \right \rangle
\label{eq:5.5}
\end{equation}
reveals a close relation with the time correlation of the corresponding 
largest perturbed shell $u_i(t)$, as shown in Fig.~{\ref{fig3}.
A measure of the relaxation time can be provided by 
the halving times $T_{1/2}$ of the mean response, at which  
$\langle \delta E_i(T_{1/2}) \rangle = 1/2 \langle \delta E_i(0) \rangle$.
The dependence of response times 
on the amplitude of the initial perturbation,
shown in Fig.{\ref{fig4}, reflects Kolmogorov scaling for 
characteristic times $\tau_n \sim k_n^{-2/3} \sim u_n^2 \sim \delta E_n$
\begin{equation}
T_{1/2} \sim \delta E \, \, .
\label{eq:5.6}
\end{equation}

The above results on the shell model show that 
the response to a finite size perturbation
of a system with many characteristic times
may depend on the amplitude of the perturbation.  
Thanks to the existence of F/R relation is possible to 
establish a link between relaxation times 
of different perturbation and characteristic times of the system. 

\section{Conclusions}
\label{sec:6}

Starting from the seminal works of Leith \cite{L75,L78},
who proposed the use of F/R relation for understanding the
response of the climatic system to changes in the external
forcing, many authors tried to apply this relation to
different geophysical problems, ranging from simplified models
\cite{Bell80}, to general circulation models \cite{North93,Kaskins97}
and to the covariance of satellite radiance spectra \cite{Kaskins99}.
In most of the applications it has not been taken into account
the limits of applicability of the F/R relation which has been used
as a kind of approximation. We have shown that a F/R relation holds
under very general conditions. The derivation in Section~\ref{sec:2}
clearly shows the limits of applicability in its simplest 
form (i.e. the Gaussian approximation (\ref{eq:2.8})).

Our main result is the demonstration that an exact fluctuation/response 
relation holds also for non infinitesimal perturbation.
This relation involves the detailed form of the
invariant probability distribution.
In particular, in order to predict the mean response to large 
perturbations, one needs a precise knowledge of the tails of the pdf.

We believe that this generalization of the usual linear response 
theory can be relevant in many applications. As an example, we can
mention climate research, where our results imply the possibility,
at least in principle, to understand the behavior of the system after a large
impulsive perturbation (e.g. a volcanic eruption) in terms of
the knowledge obtained from its time history.
Of course one has to take into account the strong limitations
due to the need to have a good statistics of rare events.

\begin{acknowledgments}
This work has partially supported by  
MIUR (Cofinanziamento {\it Fisica Statistica di Sistemi 
Complessi Classici e Quantistici}).
We acknowledge the allocation of computer resources
from INFM Progetto Calcolo Parallelo.
Angelo Vulpiani acknowledges support from the 
INFM {\it Center for Statistical Mechanics and Complexity} (SMC).
We thank M. Falcioni for useful remarks.
\end{acknowledgments}

\section{Appendix}
\label{app:1}

In this appendix we want to discuss how van Kampen criticism
is relevant for the numerical evaluation of infinitesimal
response function.

In numerical simulations, $R_{i,j}(t)$ is computed 
perturbing the variable $x_i$ at time $t=t_0$ with
a small perturbation of amplitude $\delta x_i(0)$ 
and then evaluating the separation $\delta x_i(t)$ 
between the two trajectories ${\bm x}(t)$ and ${\bm x}'(t)$
which are integrated up to a prescribed time $t_1=t_0+\Delta t$.  
At time $t=t_1$ the variable $x_i$ of the reference trajectory is again 
perturbed with the same $\delta x_i(0)$, and a new sample
$\delta {\bm x}(t)$ is computed and so forth.
The procedure is repeated $M \gg 1$ times and the mean
response is then evaluated according to (\ref{eq:2.8}).

In presence of chaos, the two trajectories ${\bm x}(t)$ and ${\bm x}'(t)$
typically separate exponentially in time 
and the perturbed system relaxes to the unperturbed one only in average, 
therefore the mean response is the result of a delicate balance of
terms which grow in time in different directions. 
The average error in the computation of $R_{i,j}(t)$
typically increases in time as $e^{L(2) t/2}/ \sqrt{M}$,
where $L(2)$ is the generalized
Lyapunov exponent \cite{BJPV98}. Thus very high
statistics is needed in order to compute $R_{i,j}(t)$ for
large $t$ \cite{F90}.

We remark that the exponential growth is generally valid only 
for infinitesimal perturbation. When the perturbation 
reaches the typical size of the system, the difference between 
the perturbed and the unperturbed trajectory tends to saturate. 
Thus, for finite amplitude perturbations the mean response is the average
of terms that remain of order $O(1)$, and
less statistics is required to obtain convergence.   
In this sense the mean response to finite perturbation 
is more representative of the behavior of a single perturbation
than in the infinitesimal case.

On the other hand, even if 
equation (\ref{eq:2.10}) is formally valid for arbitrary large 
perturbations, for practical use an upper limit exist due to 
finiteness of statistics. 
To predict the relaxation of a perturbation $\delta {\bm x}(0)$,
one needs sufficient statistics for the convergence of
$\rho ({\bm x}(0) - \delta {\bm x}(0))$.
This request is more severe in systems where large fluctuations
are suppressed.
An example is provided by the stochastic model (\ref{eq:4.1})
with
\begin{equation}
U(x)= {1 \over 2} x^2 + {1 \over 4} x^4 
\label{eq:4.4}
\end{equation}
Here the pdf has sub-Gaussian tails,
and we observe the opposite behavior of the system (\ref{eq:4.3}), 
as shown in Fig.~\ref{fig5}.
While in the case with exponential tails we have a
good statistical convergence for a perturbation greater than $2 \sigma$
in the second system this perturbation is too large to obtain
convergence even with huge statistics ($10^9$ runs). 




\begin{figure}
\centerline{\epsfxsize=8cm\epsfbox{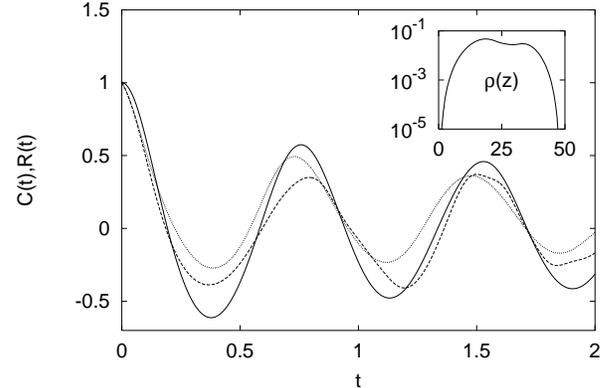}}
\caption{
Correlation function of the $z$ variable of Lorenz model (solid line)
compared with the mean response to different perturbations
of the same variable.
$\delta z_0 = 10^{-2} \sigma$ (dashed line),
$\delta z_0 = \sigma$ (dotted line),
with
$\sigma = \sqrt{ \langle z^2 \rangle - {\langle z \rangle}^2} = 8.67$.}
\label{fig1}
\end{figure}


\begin{figure}
\centerline{\epsfxsize=8cm\epsfbox{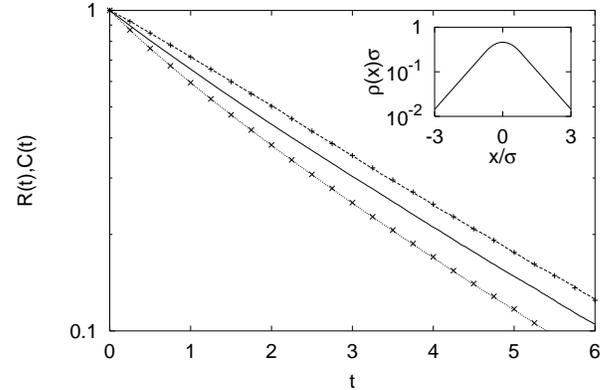}}
\caption{Mean response of the stochastic differential equation
$dx/dt= -B(x) + \sqrt{2D} \xi(t)$,
with $D=1$, $B(x) = x$ for $|x| < 1$ 
and $B(x) = 1$ for $|x| > 1$,
to different perturbations: 
large $\delta x_0 = 2.3 \sigma$ ($+$) and 
infinitesimal $\delta x_0 = 7.6 \times 10^{-3} \sigma $ ($\times$). 
In both cases the mean response is exactly predicted by the correlator
$<x(t)f({\bm x}(0))>$
(dashed line for $\delta x_0 = 2.3 \sigma$ 
and dotted line for $\delta x_0 = 7.6 \times 10^{-3}$)  
according to Eq.(\ref{eq:2.10})
while the simple correlation $\langle x(t)x(0) \rangle / \sigma^2$
(solid line) just gives an estimate of the relaxation time. 
In the inset we show the invariant probability distribution 
$\rho(x)\sigma$ versus $x/\sigma$ with 
$\sigma = \sqrt{ \langle x^2 \rangle - {\langle x \rangle}^2} = 1.32$.
Statistics is over $10^6$ independent runs.  
}
\label{fig2}
\end{figure}


\begin{figure}
\centerline{\epsfxsize=8cm\epsfbox{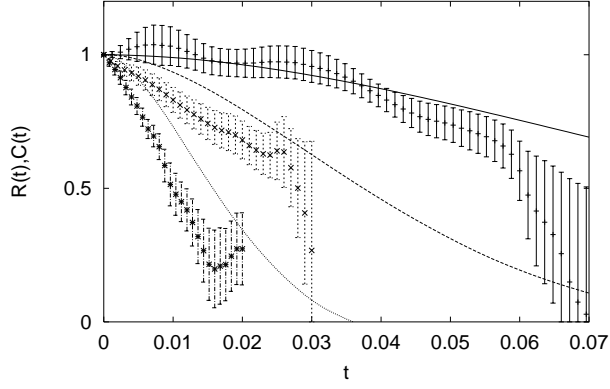}}
\caption{Mean response
$R(t)= \langle \delta E(t) / \delta E(0) \rangle$
of the total energy $E(t) = 1/2 \sum |u_n|^2$
of the shell model (\ref{eq:5.1})
to different amplitude perturbations:
$\delta E(0) =5.5 \times 10^{-3}$ ($+$),
$\delta E(0) =1.7 \times 10^{-3}$ ($\times$),
$\delta E(0) =4.5 \times 10^{-4} $ ($*$). 
Varying the amplitude of the initial perturbation
different relaxation rates are observed, and the response function 
is roughly similar to the 
correlation function of the corresponding largest perturbed shell:
shell $n = 12$ (solid line), 
shell $n = 14$ (dashed line),
shell $n = 16$ (dotted line).}
\label{fig3}
\end{figure}


\begin{figure}
\centerline{\epsfxsize=8cm\epsfbox{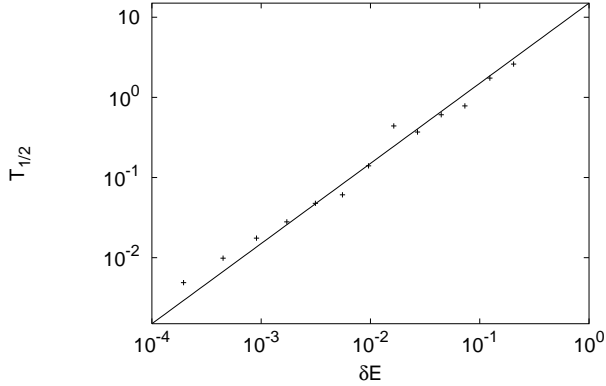}}
\caption{Halving times $T_{1/2}$ of the mean response 
to different amplitude perturbations 
of the total energy $E(t) = {1/2} \sum |u_n|^2$ 
of the shell model (\ref{eq:5.1}):
$ R(t)= \langle \delta E(t) / \delta E(0) \rangle$.
Solid line represents the dimensional scaling 
$T_{1/2} \sim \delta E$.}
\label{fig4}
\end{figure}


\begin{figure}
\centerline{\epsfxsize=8cm\epsfbox{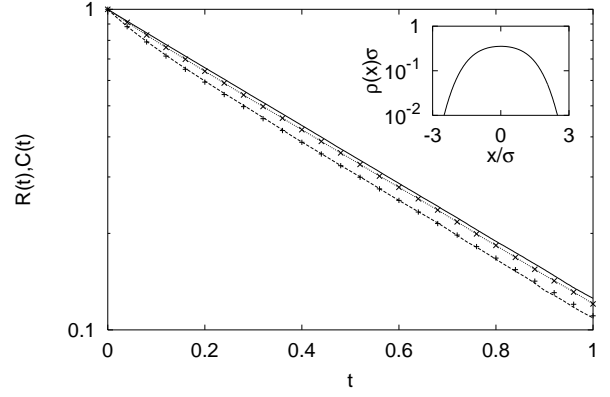}}
\caption{Mean response of the stochastic differential equation
$dx/dt = -B(x) + \sqrt{2D} \xi(t)$,
with $D=1$, $B(x) = x+x^3$,
to different perturbations:
finite $\delta x_0 = 1.5 \sigma$ ($+$) and
infinitesimal $\delta x_0 = 1.5 \times 10^{-2} \sigma$ ($\times$).
The mean response is exactly predicted by the correlator 
$<x(t)f({\bm x}(0))>$
(dashed line for $\delta x_0 = 1.5 \sigma$ 
and dotted line for $\delta x_0 = 1.5 \times 10^{-2} \sigma$)  
according to Eq.(\ref{eq:2.10}), 
while the simple correlation $\langle x(t)x(0) \rangle / \sigma^2$
(solid line) just gives an estimate of the relaxation time. 
In the inset we show the invariant probability distribution 
$\rho(x) \sigma$ versus $x/\sigma$ with 
$\sigma = \sqrt{ \langle x^2 \rangle - {\langle x \rangle}^2} = 0.68$.
Statistics is over $10^6$ independent runs.}
\label{fig5}
\end{figure}

\end{document}